\pdfoutput=1
\documentclass[aps,prl,reprint,amsmath,amssymb,superscriptaddress]{revtex4-1}
\usepackage{bm}
\usepackage{graphicx}
\usepackage[colorlinks]{hyperref}
\usepackage{mathrsfs}
\usepackage{times}

\newcommand{\beq}{\begin{equation}}
\newcommand{\eeq}{\end{equation}}

\begin{document}

\title{Bosonic superfluid on lowest Landau level}

\author{Sergej Moroz}
\affiliation{Department of Physics, Technical University of Munich, 85748 Garching, Germany}
\affiliation{Munich Center for Quantum Science and Technology (MCQST), Schellingstr. 4, D-80799 M\"unchen}

\author{Dam Thanh Son}
\affiliation{Kadanoff Center for Theoretical Physics, University of Chicago, Chicago, IL 60637, USA}

\begin{abstract}
We develop a low-energy effective field theory of a two-dimensional bosonic superfluid on the lowest Landau level at zero temperature and identify a Berry term that governs the dynamics of coarse-grained superfluid degrees of freedom. For an infinite vortex crystal we compute how the Berry term affects the low-energy spectrum of soft collective Tkachenko oscillations and non-dissipative Hall responses of the particle number current and stress tensor. This term gives rise to a quadratic in momentum term in the Hall conductivity, but does not generate a non-dissipative Hall viscosity.
\end{abstract}

\maketitle


\emph{Introduction.}---It is known since 1930 \cite{Landau1930}  that the quantum-mechanical energy spectrum of a nonrelativistic particle of mass $m$ and unit charge in a constant magnetic field $B$ form an infinite equidistant tower of highly degenerate Landau levels separated from each other by the cyclotron energy gap $\omega_c= B/m$
\footnote{In this paper we set $\hbar=1$ and absorb $e$ into $B$.}.
For electrons this observation, together with the Pauli exclusion principle, allows one to understand salient  features of integer quantum Hall fluids. Theoretically, in the limit of an infinite cyclotron gap, one is left only with states occupying the lowest Landau level (LLL). The LLL  Hilbert space is a natural starting point for formulating theories of strongly correlated fermionic fractional quantum Hall liquids.

On the other hand, bosons respond to an external magnetic field by
forming quantum vortices, whose density is fixed by the magnetic
field, $n_v=B/2\pi$.  At zero temperature, the phase diagram of a
homogeneous many-boson system with an isotropic short-range repulsive
interaction is determined by the ratio of the density of bosons $n$ to
the density of vortices $n_v$, known as the filling fraction
$\nu=n/n_v$. For $\nu\gtrsim\nu_{cr}\sim 8$ \cite{Cooper2001,
Sinova2002} bosons form a gapless superfluid vortex crystal, while for
$\nu\lesssim\nu_{cr}$ the system becomes an incompressible bosonic
fractional quantum Hall state, whose nature depends on the precise
filling fraction~\cite{Cooper2008, Viefers2008}.  Within the
crystalline phase, one distinguishes the regime of
slow \cite{tkachenko1965, tkachenko1966, tkachenko1969, Sonin2014} and
fast \cite{Ho2001, Sinova2002, Watanabe2004, Sonin2005,
Aftalion2005, Cozzini2006, Baym2007} rotation, with $ n_v \xi^2 \ll 1$ or $n_v \xi^2 \gg 1$ correspondingly, where $\xi$ is the superfluid coherence length \cite{Baym2003, Sonin2016}.
In the latter regime, an arbitrary number of bosons is accommodated by
a single orbital, and the LLL approximation can be applied in the regime $\nu> 1$, in contrast to the fermionic case.
In fact, it is known since the seminal work of Abrikosov \cite{Abrikosov1957} that the physics of the vortex lattice in superconductors can be understood analytically within the LLL approximation in the bosonic Ginzburg-Landau theory. 

In this Letter, we formulate and analyze a low-energy effective field theory of a two-dimensional bosonic superfluid in the LLL regime. We discover that the existence of a smooth LLL limit
implies that in this regime the dynamics of superfluid degrees of freedom is governed by what we will term ``the Berry Lagrangian''
\beq \label{LLLt}
\mathcal{L}_{B}=\mp \frac 1 2 n \varepsilon^{ij}\partial_i v_j,
\eeq
where $n$ and $v_i$ denote the smooth superfluid density and velocity, coarse-grained over regions containing a large number of quantized vortices. 
As characteristic to a Berry term, its contribution to the action depends on the
trajectories of the fluid particles, but not on the time it takes for these
trajectories to be traversed.
The sign of the Berry term \eqref{LLLt} is fixed by the direction of the external magnetic field $B$. The term \eqref{LLLt} is odd under separate time-reversal $T$ ($t\to -t$) and two-dimensional parity $P$ ($x\leftrightarrow y$) transformations.

To the best of our knowledge, the term \eqref{LLLt} did not appear in previous studies of superfluids in the LLL regime.
In this Letter, we focus our attention on the implications of the Berry term \eqref{LLLt} for the low-energy physics of an infinite unpinned vortex crystal at zero temperature. We determine how this term modifies the dispersion of the collective Tkachenko wave (see Eq. \eqref{omega2k4k6}) and demonstrate that it gives rise to a quadratic in momentum term in the Hall conductivity (see Eq. \eqref{Hcond}). In addition, we find that this term generates a time-reversal odd contribution to the stress tensor \eqref{STB}, but does not produce a non-dissipative Hall viscosity.

Our derivation of the Berry term is quite general and suggests that
such a term should also appear
away from the LLL limit, which is deferred to a future work. On the other hand, our conclusions are not directly applicable in the incompressible strongly-correlated non-superfluid regime.



\emph{Effective field theory of LLL superfluid.}---Our starting point is the microscopic theory of an interacting nonrelativistic spinless bosonic field $\psi$ coupled to a $U(1)$ gauge field $A_{\mu}$ and, for later convenience, an external spatial metric $g_{ij}$. The Lagrangian is given by
\beq \label{micth}
\mathcal{L}=\frac i 2\psi^{\dagger}\overleftrightarrow D_t \psi-\frac{g^{ij}}{2m}D_i \psi^\dagger D_j \psi+\frac{\text{g} B}{4m}\psi^\dagger \psi+ \mathcal L_{int}(n),
\eeq
where $D_\mu \psi=(\partial_\mu-iA_\mu)\psi$, $\text{g}$ denotes the gyromagnetic factor of elementary bosons and $B=\varepsilon^{ij}\partial_i A_j$ is the magnetic field \footnote{The Levi-Civita tensor is defined as $\varepsilon^{ij}=g^{-1/2} \epsilon^{ij}$, where $g=\det g_{ij}$ and $\epsilon^{12}=1$.}.  We assume that the interaction Lagrangian depends only on the density $n=\psi^\dagger \psi$.

First, we will work in flat space and introduce the complex spatial coordinates
$z=x+iy$ and $\bar z=x-iy$. In these coordinates, the free part of the microscopic Lagrangian is
\beq \label{Lmic}
\mathcal{L}_0=\frac i 2\psi^{\dagger}\overleftrightarrow D_t \psi-\frac{2}{m}  D_z \psi^{\dagger}D_{\bar{z}} \psi+(\text{g}-2)\frac{B}{4m} \psi^\dagger \psi.
\eeq
In the symmetric gauge a constant background magnetic field $B_0$ results from $A_z=-i B_0 \bar z/4$ and $A_{\bar z}=i B_0 z/4$. We notice that $\text{g}=2$ is special since in this case the last term drops out and the energy of the LLL vanishes. In fact in this case the LLL is degenerate even in inhomogeneous magnetic field on arbitrary curved surface \cite{Aharonov1979, Alicki1993}. We can project the Lagrangian \eqref{Lmic} to the LLL by taking the limit $m\to 0$, while keeping the magnetic field $B$ fixed resulting in a diverging cyclotron frequency $\omega_c$. Using the Hubbard-Stratonovich transformation, we rewrite now Eq. \eqref{Lmic} as
\beq \label{Lchi}
\mathcal{L}_0=\frac i 2 \psi^{\dagger}\overleftrightarrow{D_t} \psi-\chi^{\dagger}D_{\bar{z}} \psi-\chi D_{z} \psi^{\dagger}+\frac m 2 \chi^\dagger \chi+(\text{g}-2)\frac{B}{4m} \psi^\dagger \psi.
\eeq
In the limit $m\to 0$, the equation of motion for the Lagrange multiplier $\chi^\dagger$ produces the holomorphic condition $D_{\bar z} \psi=0$, which is solved by a LLL wave function $\psi_{LLL}(z,\bar z)=f(z) e^{-|z|^2/(4 l_B^2)}$, where we introduced the magnetic length $l_B=\sqrt{1/B_0}$. Notably in this construction the LLL limit is smooth only for $\text{g}=2$. While for Dirac electrons $\text{g}=2$ is the physical value of the gyromagnetic factor, for bosons $\text{g}=2$ is not a natural choice. For example, neutral spinless bosons rotating with angular frequency $\Omega=B/2m$ couple to the $U(1)$ source minimally and thus in that case $\text{g}=0$.
Fortunately, from Eq. \eqref{micth} one readily observes that  given the Lagrangian for $\text{g}=2$ fixes the theory for any value of $\text{g}$
\beq \label{shift}
\mathcal{L}^\text{g}(\dots, A_0)=\mathcal{L}^{\text{g}=2}(\dots, A_0+\frac{\text{g}-2}{4m}B).
\eeq
So our strategy in the following is to construct the low-energy effective theory first for $\text{g}=2$, making use of the smoothness of the LLL, and subsequently transform the parameter $\text{g}$ to its physical value using Eq. \eqref{shift}.

In this Letter we use boson-vortex duality \cite{peskin1978, Dasgupta1981} to derive the low-energy effective theory of a superfluid on LLL. In this formulation superfluid degrees of freedom are encoded in a $2+1$-dimensional gauge field $a_{\mu}$, while quantum vortices are charged particles with respect to this field. In this language, a vortex crystal is a two-dimensional Wigner solid of dual point charges embedded in a static neutralizing background and a homogenous dual magnetic field fixed by the superfluid density \cite{MHBS2018}.
The superfluid part of the LLL effective theory can be derived from the microscopic model \eqref{Lchi} as follows: First, we parametrize the bosonic field $\psi=\sqrt{n}e^{i\theta}$, where $\theta$ is the phase bosonic field. Averaging now over regions with large number of vortices, such that their total charge is fully compensated by the charge density of a static neutralizing background, the dual gauge Lagrangian is defined as the Legendre transformation $\tilde {\mathcal{L}}_0=\mathcal{L}_0-\pi^{\mu} \partial_\mu \theta
$, where $\pi^\mu=\partial \mathcal{L}_0/\partial \partial_\mu \theta$. In the the LLL limit $m=0$ and $\text{g}=2$ one finds \cite{supmat}
\beq \label{dualg2}
\tilde{\mathcal{L}}_{0}^{LLL, \text{g}=2}
=\varepsilon^{\mu\nu\rho} A_\mu \partial_{\nu} a_{\rho}+\frac{1}{2b} e^i \partial_i b,
\eeq
where the dual magnetic and electric fields $b=\varepsilon^{ij}\partial_i a_j$ and $e_i=\partial_t a_i-\partial_i a_t$ were defined in terms of the $U(1)$ particle number current 
$
j^{\mu}=\delta S/\delta A_\mu=\varepsilon^{\mu\nu\rho}\partial_\nu a_\rho
$
. 
It is straightforward to check that the change of the sign of the background magnetic field $B_0\to -B_0$ modifies the Lagrangian \eqref{Lchi} resulting in the anti-holomorphic LLL condition $D_z \psi=0$ which in turn reverses the sign of the second term in the dual Lagrangian \eqref{dualg2}. By introducing now the superfluid velocity 
\beq \label{vel}
v^i=j^i/n=-\varepsilon^{ij}e_j/b
\eeq
 and performing integration by parts, we can cast this term into the form \eqref{LLLt}. In the gauge theory language the Berry term is linear in the electric field and thus does not affect the Hamiltonian. Due to this term, however, the Poisson brackets in the LLL regime should differ from the canonical structure of the Hamiltonian theory in the Tkachenko regime derived in Refs.~\cite{volovik1979, volovik1980}.

It straightforward to generalize the above duality transformation to the case with $m\ne 0$ and $\text{g}\ne 2$. We start from Eq. \eqref{Lchi} and follow the steps above
\beq \label{dualgg}
\tilde{\mathcal{L}}_0=\frac{m e^2}{2b}+\tilde{\mathcal{L}}_{0}^{LLL, \text{g}=2}+\frac{(\text{g}-2)}{4m} B b.
\eeq 
Hence the coefficient of the Berry term does not depend on the values of $\text{g}$ and $m$. In the LLL limit the electric term $m e^2/(2b)$ drops out and the dynamics of the superfluid is governed only by the Berry term.

Alternatively, one can derive the effective theory \eqref{dualgg} from the general coordinate (diffeomorphism) invariance which is inherited from the microscopic model \eqref{micth}. In the dual theory the coupling of bosons to the $U(1)$ source is encoded in the mixed Chern-Simons (CS) action $S_{CS}=\int\! A d a$. It was shown however in Ref.~\cite{Son2013NC} that for $m \ne 0$ and $\text{g}\ne 0$ the gauge field $A_{\mu}$ does not transform as a one-form under time-dependent spatial diffeomorphisms and thus on its own the mixed CS action is not general coordinate invariant. To circumvent this problem in the theory with $\text{g}=2$ we introduce the improved gauge potential $\tilde A_\mu$ that transforms as a one-form \cite{Son2013NC}
\beq
\begin{split}
&A_0 \to \tilde A_0= A_0-\frac m 2 g_{ij} v^i v^j-\frac 1 2 \varepsilon^{ij}\partial_i v_j, \\
&A_i\to \tilde A_i= A_i+m v_i,
\end{split}
\eeq
where the velocity field was defined in Eq. \eqref{vel}. The improved mixed CS term is diffeo-invariant and reads
\beq
\begin{split}
\mathcal{L}_{\tilde{CS},\text{g}=2}&=\tilde A d a=A d a+\frac{m g^{ij} e_i e_j}{2b}-\frac 1 2 b \nabla_i \left( \frac{e^i}{b}\right),
\end{split}
\eeq
where $\nabla_i$ is the Levi-Civita covariant derivative. To get the diffeo-invariant mixed CS term in the theory with $\text{g}\ne 2$ we use Eq. \eqref{shift}.
In flat space the end result reproduces Eq. \eqref{dualgg} up to a surface term.

How does the interaction term $\mathcal{L}_{int}$ affect the dual theory? As long as it depends only on the density, it is general coordinate invariant on its own and its role in the duality transformation is completely passive. The Lagrangian of the dual theory with this type of interactions is $\tilde {\mathcal{L}}=\tilde {\mathcal{L}}_0+\mathcal{L}_{int}(b)$.

\emph{Angular momentum on LLL.}---%
Taking the derivative of the dual effective action  with respect to $A_i$ one finds the particle number current
\beq \label{cur}
j^i=\frac{\delta S}{\delta A_i}=-\varepsilon^{ij} e_j+\frac{\text{g}-2}{4m}\varepsilon^{ij} \partial_j b=n v^i+\frac{\text{g}-2}{4m}\varepsilon^{ij} \partial_j n.
\eeq
In a Galilean-invariant system the current $j^i$ fixes uniquely the momentum density \cite{Geracie2015}
\beq \label{momentum}
T^{0i}= m j^i-\frac{\text{g}}{4}\varepsilon^{ij}\partial_j n.
\eeq
Substituting now Eq. \eqref{cur} into this equation and taking the limit $m\to 0$, the angular momentum in the LLL superfluid is $J_z=\int d^2x \epsilon_{ij} x^i T^{0j}=-N$, where $N$ is the total number of elementary bosons. Thus we recover the known result that the angular momentum of the LLL superfluid is given in absolute value by the number of particles~\cite{supmat}. In the studies of rotating neutral superfluids in the LLL approximation, it is interpreted as a universal deviation of the angular momentum in the lab frame from the solid-body prediction (see Ref.~\cite{ Fetter2009} and references therein).


\emph{Infinite vortex lattice in the LLL regime.}---Now we are ready to include the physics of vortices. In an infinite two-dimensional system in a magnetic field quantum vortices in a superfluid form a triangular Abrikosov lattice. The effective Lagrangian of the vortex degrees of freedom used here is inspired by Ref.~\cite{MHBS2018}, where the dual effective theory of a vortex crystal (at $\text{g}=0$) was constructed away from the LLL limit.   The effective theory of an infinite vortex crystal analyzed in this paper is \footnote{Importantly, in Ref. \cite{MHBS2018} $j^i=-\varepsilon^{ij} e_j$ which differs from the relation \eqref{cur}. As a result, the Berry term is included implicitly in Ref.~\cite{MHBS2018} within the $me^2/(2b)$ term.  In addition, in Ref. \cite{MHBS2018} the sign convention for the $U(1)$ source $\mathcal{A}_\mu$ was chosen opposite.} 
\begin{multline}\label{EFTflat}
\mathcal{L}_{vc}=\frac{m e^2}{2b}+\frac{1}{2b} e^i \partial_i b + \epsilon^{\mu\nu\rho}  \mathcal{A}_\mu \partial_\nu a_\rho +\frac{(\text{g}-2)}{4m} \mathcal{B} b  -\varepsilon(b)\\
-\frac{B_0}{2} b \epsilon_{ij}u^i  D_t u^j+ B_0 e_i u^i- \mathcal{E}_{\text{el}}(u_{ij}),
\end{multline}
where $\varepsilon(b)$ is the internal energy density which is fixed by the form of the interaction Lagrangian $\mathcal{L}_{int}$, $u^i$ denotes the coarse-grained displacement of vortices from their equilibrium positions and $u_{ij}=\partial_{(i} u_{j)}-\partial^k u_i \partial_k u_j/2$ is the strain tensor.
The first line describes the superfluid sector discussed above. The only difference is that here we introduced a modified $U(1)$ source. The temporal component  $\mathcal{A}_0=A_0+\frac{\text{g}-2}{4m}B_0$ measures the chemical potential from the LLL, while the spatial part  $\mathcal{A}_i$ is set to vanish in the ground state. The second line encodes the physics of vortices: the Magnus term governs the dynamics, the second term measures the dipole energy density and the last term defines the elastic properties of the triangular lattice which to lowest order in derivatives coincide with the elastic energy density of an isotropic medium and thus depend only on the compressional and shear bulk moduli $C_1$ and $C_2$ (see Refs.~\cite{Baym2003, MHBS2018} for details). Expanding now around $e_i=0$ and the minimum $b=n_0$ of $\varepsilon(b)$, to quadratic order the Lagrangian in the absence of the $U(1)$ source is
\begin{multline} \label{LqLLL}
\mathcal{ L}_{vc}^{(2)}=-\frac{B_0 n_0}{2} \epsilon_{ij}u^i\dot{u}^j+B_0 e_i u^i-\mathcal{E}^{(2)}_{\text{el}}(\partial u)-\frac{\varepsilon''}{2}\delta b^2 \\
+\frac{m e^2}{2n_0}+\frac{1}{2n_0} e^i \partial_i \delta b, 
\end{multline}
where $\delta b=b-n_0$ and the speed of sound $c_s$ is fixed by $ \varepsilon''=m c_s^2/n_0 $.
In the power-counting defined in Ref.~\cite{MHBS2018} the first line defines the leading order (LO) Lagrangian which gives rise to a transverse Tkachenko wave with a gapless quadratic dispersion relation \cite{Sonin1976, volovik1979, Baym2003}. The electric and the Berry terms are the next-to-leading order (NLO) corrections within this power-counting scheme.
Here we investigate the LLL regime $m\to 0$, where the electric term drops out. In addition, the speed of sound is scaled to infinity such that $ \varepsilon''=m c_s^2/n_0$ remains constant. While one can write a number of additional NLO terms that preserve $P$ and $T$ separately, we are not aware of other NLO $P$- and $T$-odd terms allowed in this system except for the phonon Hall viscosity introduced in Ref.~\cite{Barkeshli2012}. The fate of the phonon Hall viscosity term in the LLL regime is unclear now and it is not discussed here.  

By solving the equations of motion for a plane wave propagating along say $x$-axis we find the elliptically-polarized gapless Tkachenko mode with the dispersion ($k=-i \partial_x$, $\omega=i\partial_t$)
\beq\label{omega2k4k6}
\begin{split}
\omega^2=\frac{2 C_2 \varepsilon''
   }{B_0^2} k^4-\frac{2 C_2 \varepsilon''
   }{B_0^3}k^6+O\big(k^8\big).
\end{split}
\eeq
Due to the NLO Berry term, the dispersion starts to deviate from the quadratic form at momenta  of order of the inverse of the magnetic length,  $k\sim l_B^{-1}$.
This correction is absent in the results of Baym (Eq.~(14) in Ref.~\cite{Baym2003}) and Sonin (Eq.~(3.128) in Ref.~\cite{Sonin2016}),  which give a strictly quadratic dispersion in the LLL limit defined above. 
Note that the Lagrangian~(\ref{EFTflat}) contains only the minimum number of
terms required by Galilean invariance and the regularity of the LLL limit;
additional terms may change the coefficient of the $k^6$ term in Eq.~({\ref{omega2k4k6}). In \cite{supmat} we compare the equations of motion derived from the Lagrangian \eqref{LqLLL} with the hydrodynamic equations found in Ref.~\cite{Baym2003}. 

By using the Lagrange coordinates $X^a(t, \mathbf{x})$ frozen into the vortex lattice instead of displacements $u^i(t,\mathbf{x})$, the effective theory of the vortex crystal can be cast into the general coordinate invariant form \cite{MHBS2018}. In this formulation the effective action of the vortex crystal is $S_{vc}=\int dt d^2x \sqrt{g} \mathcal{L}_{vc}$ with
\begin{multline} \label{EFTcurve}
\mathcal{L}_{vc}=\frac{m g^{ij} e_i e_j}{2b}+\frac {1} {2b}  g^{ij} \nabla_i  b   e_j + \epsilon^{\mu\nu\rho}  A_\mu \partial_\nu a_\rho +\frac{(\text{g}-2)}{4m} B b  \\
-\varepsilon(b) -\pi n_v \varepsilon^{\mu\nu\rho} \epsilon_{ab} a_\mu \partial_\nu X^a \partial_\rho X^b-\mathcal{E}_{\text{el}}(U^{ab}),
\end{multline}
where $U^{ab}=g^{ij}\partial_i X^a \partial_j X^b$ with $a=1,2$. Remarkably, the $P$- and $T$-odd Magnus and dipole terms combine into a single topological term. Here we focus on the $P$- and $T$-odd Berry term which depends on the metric and thus modifies the stress tensor $T_{B}^{ij}=2 \delta S_{B}/(\sqrt{g}\delta g_{ij})$
\begin{multline} \label{STB}
T_{B}^{ij}=\frac 1 2  g^{ij} \nabla^k (\log b) e_k- \nabla^{(i} (\log b) e^{j)} +\frac 1 2 g^{ij}g^{kl}\nabla_k e_l \\
 =-\frac 1 2  \big( g^{ik} \varepsilon^{jl}  +g^{jk} \varepsilon^{il}    \big) v_k \nabla_l n  +\frac n 2 g^{ij}\varepsilon^{kl} \nabla_k  v_l,
\end{multline}
where the second line was obtained by using the duality dictionary and the two-dimensional vector identity $ \big( g^{ik} \varepsilon^{jl}  +g^{jk} \varepsilon^{il}    \big) a_{[k}b_{l]}=g^{ij} \varepsilon^{kl} a_k b_l
$. Since in the LLL limit at $\text{g}=2$ the momentum density \eqref{momentum} reduces to $T^{0i}=-\varepsilon^{ij}\partial_j n/2$, it is straightforward to check that the Berry stress tensor \eqref{STB} transforms properly under Galilean boosts \cite{supmat}. In the case $\text{g}\ne 2$, the last term on the first line of Eq. \eqref{EFTcurve} gives a modification to the pressure $\Delta T_{LLL}^{ij}=-(\text{g}-2) B b g^{ij}/(4m)$ which is consistent with general relations derived in Appendix B of Ref.~\cite{Geracie2015}.


\emph{Hall responses.}---Being odd under $P$ and $T$, the Berry term affects the non-dissipative Hall response of the vortex crystal. Here we discuss linear Hall response of the particle current and stress tensor in the LLL regime. We mostly focus  on the case $\text{g}=2$, where all response functions are regular. The case of $\text{g}\ne 2$ is discussed in detail in \cite{supmat}.

Given our definition of the $U(1)$ electric field $E_i=\partial_t A_i-\partial_i A_t$, the conductivity tensor is defined as
\beq \label{cond}
\sigma^{ij}(\omega, \mathbf{k})=-\frac{\delta j^i}{\delta E_j}=\frac{\delta j^i}{i \omega \delta A_j}=\frac{\delta j^i}{i k_j \delta A_t} \,.
\eeq
Substituting now linearized equations of motion derived from the Lagrangian \eqref{EFTflat} into the definition of the current \eqref{cur}, we find with the help of Eq. \eqref{cond} in the LLL $m\to 0$ limit the Hall conductivity $\sigma^H=(\sigma^{xy}-\sigma^{yx})/2$
\beq \label{Hcond}
\sigma^{H}(\omega, k)=\frac{n_0}{B_0}-\frac{n_0}{2 B_0^2}k^2+O(k^4).
\eeq
The quadratic term in momentum originates from the NLO Berry Lagrangian.

The geometric response of the stress tensor to the metric allows one to define the elasticity and viscosity tensors $\lambda^{ijkl}$ and $\eta^{ijkl}$
\beq
\delta T^{ij}=-\lambda^{ijkl} \delta g_{kl}-\eta^{ijkl} \delta \dot g_{kl}.
\eeq
The Hall viscosity tensor $\eta^{ijkl}_H$ \cite{Avron1995, Avron1997} is defined as the odd part of the viscosity tensor under $ij\leftrightarrow kl$
\beq
\eta_H^{ijkl}=\frac 1 2 (\eta^{ijkl}-\eta^{klij}).
\eeq
This non-dissipative response is ubiquitous in two-dimensional systems which break time-reversal symmetry \cite{Avron1995, Avron1997, Read2009, Read2011}. In a rotation-invariant system the Hall viscosity has only one independent component \cite{Avron1995, Avron1997} which we denote as $\eta_H$.
To compute this response function we first derive from the Lagrangian \eqref{EFTcurve} the equations of motion in the LLL regime $m=0$ and at $\text{g}=2$
\beq \label{EOM1}
-\frac {1} {2} g^{ij} \nabla_i \nabla_j \log b+\pi n_v \varepsilon^{ij}\epsilon_{ab} \partial_i X^a \partial_j X^b-B=0,
\eeq
\beq \label{EOM2}
\begin{split}
&\frac {1} {2\sqrt{g}} \varepsilon_{jk} \partial_t \Big(\sqrt{g} g^{kl} \nabla_l \log b \Big)-\frac{1}{2}\partial_j \frac{g^{kl}\nabla_k e_l}{b} \\
&-\varepsilon''(b) \partial_j b+2\pi n_v \epsilon_{ab} \dot X^a \partial_j X^b-E_j=0,
\end{split}
\eeq
\beq \label{EOM3}
\pi n_v \varepsilon^{\mu\nu\rho} \epsilon_{ab} \partial_\mu a_\nu \partial_\rho X^b-\frac{1}{\sqrt{g}}\partial_j \big(\sqrt{g} \frac{\partial \mathcal{E}_{\text{el}}}{\partial U^{ab}} g^{ij}\partial_j X^b \big)=0
\eeq
and linearize them around the flat space solution $b=n_0$, $e_i=0$ and $X^a=\delta^a_i x^i$. To extract the Hall viscosity it is sufficient to restrict the perturbation $\delta g_{ij}(t)$ to be traceless and homogenous in space. An explicit calculation shows that  such a perturbation $\delta g_{ij}(t)$ does not affect linearized equations of motion. This result implies that the variation of the stress tensor, derived from the Lagrangian \eqref{EFTcurve}, with respect to the time derivative of the homogeneous traceless metric is zero and thus the Hall viscosity must vanish. This conclusion does not change if one generalizes the calculation to the case with $\text{g}\ne 2$. 

Galilean invariance gives rise to relations between the conductivities and viscoelastic response functions \cite{Hoyos2012, Bradlyn2012}. In the LLL limit $m\to0$ and at $\text{g}=2$, where all conductivities are regular, the Hall viscosity is completely fixed by the Hall conductivity \cite{Geracie2015}
\beq \label{condvis}
\eta_{H}=\frac{1}{2}B_0^2 \partial^2_{k}\sigma_H(k)+\frac{1}{2}B_0 \sigma_H(k=0).
\eeq
After substituting into this relation the result \eqref{Hcond}, we find $\eta_{H}=0$. For $\text{g}\ne 2$ the LLL limit is not smooth which leads to singular terms in the conductivity tensor. As a result, the simple relation \eqref{condvis} is not applicable anymore and must be replaced by a more complicated relation that involves the singular part of the longitudinal conductivity, for a detailed discussion we refer to \cite{supmat}. Nevertheless, the Hall viscosity derived from that relation is still zero which agrees with the result of the geometric calculation.

In summary, the $P$- and $T$-odd NLO Berry term fixes in the LLL regime the quadratic in momentum term in the Hall conductivity, but does not give rise to the dissipationless Hall viscosity.


\emph{Conclusions.}---%
In a two-dimensional bosonic superfluid the existence of a smooth LLL limit allowed us to identify a Berry term in the low-energy effective theory. Since the coefficient of this Berry term does not depend on the mass of the elementary boson it appears that it survives in a bosonic superfluid in magnetic field even away from the LLL limit. As a result, in the effective theory developed here the crossover of a vortex crystal  from the LLL to the Tkachenko regime is controlled by the mass of the boson $m$. While in the LLL limit the dynamics of the coarse-grained superfluid is governed by the Berry term $e^i \partial_i b/2b$, in the Tkachenko limit $m\to \infty$ the electric term $me^2/2b$ dominates.

The Berry term can be interpreted as the coupling of the dynamical electric field $e_i$ to the dipole density $d^i= \partial^i \log b/2$. From Eq. \eqref{momentum}, in the LLL limit $m\to 0$ and at $\text{g}=2$ the dipole density is
$
d_i=\varepsilon_{ij} T^{0j}/b,
$
i.e., proportional and perpendicular to the momentum density $T^{0i}$.

In the future it would be interesting to determine the fate of the phonon Hall viscosity term \cite{Barkeshli2012} in the LLL regime and find if the theory developed here is related to hydrodynamics of a vortex fluid in an incompressible liquid \cite{Wiegmann2014}. It would be also useful to understand the role of the Berry term in vortex crystals in two-dimensional fermionic chiral $p+ip$ superfluids, where Cooper pairs have the gyromagnetic factor $\text{g}=2$ and the geometric spin $s= 1$ \cite{Read2011, Hoyos2013, Moroz2014a}.


\emph{Acknowledgements.}---We acknowledge useful discussions with Carlos Hoyos, Omri Golan, Eduoard Sonin and Wilhelm Zwerger.  The work of S.M.~is funded by the Deutsche Forschungsgemeinschaft (DFG, German Research Foundation) under Emmy Noether Programme grant no.~MO 3013/1-1 and under Germany's Excellence Strategy - EXC-2111 - 390814868.  The work of D.T.S. is supported, in part, by Department Of Energy grant No.\ DE-FG02-13ER41958, the Army Research Office Multidisciplinary University Research Initiative grant No.\ 63834-PH-MUR, and a Simons Investigator Grant from the Simons Foundation.


\bibliography{library}


\clearpage
\begin{widetext}
\setcounter{page}{1}
\renewcommand{\theequation}{S\arabic{equation}}
\setcounter{equation}{0}

\section{Supplemental material: Bosonic superfluid on lowest Landau level}


\subsection{Duality transformation in the LLL regime at $\text{g}=2$}
We start from the free Lagrangian in the LLL regime $m\to 0$ at $\text{g}=2$
\beq
\begin{split}
\mathcal{L}_0&=\frac i 2 \psi^{\dagger}\overleftrightarrow{D_t} \psi-\chi^{\dagger}D_{\bar{z}} \psi-\chi D_{z} \psi^{\dagger} \\
&=-n (\partial_t \theta-A_t)-\Big[\frac{e^{i\theta}}{2\sqrt{n}}\chi^{\dagger} \big(\partial_{\bar z} n +2in (\partial_{\bar z} \theta- A_{\bar z}) \big)+H.c.\Big].
\end{split}
\eeq
The dual Lagrangian is given by
$
\tilde {\mathcal{L}}_{0}=\mathcal{L}_0-\pi^{\mu} \partial_\mu \theta
$, where $\pi^\mu=\partial \mathcal{L}_0/\partial \partial_\mu \theta$. Using
\beq
\begin{split}
\pi^t&=\frac{\partial \mathcal{L}_0}{\partial \partial_t \theta}=-n, \\
\pi^{z}&=\frac{\partial \mathcal{L}_0}{\partial \partial_{ z} \theta}=i e^{-i\theta}\sqrt{n} \chi, \\
\pi^{\bar z}&=\frac{\partial \mathcal{L}_0}{\partial \partial_{\bar z} \theta}=-i e^{i\theta}\sqrt{n} \chi^{\dagger} \\
\end{split}
\eeq
we find
\beq
\tilde{\mathcal{L}}_0=n A_t-\Big[\frac{e^{i\theta}}{2\sqrt{n}}\chi^{\dagger} \big(\partial_{\bar z} n -2in  A_{\bar z} \big)+H.c.\Big].
\eeq
Now using
\beq \label{currs}
\begin{split}
j^0&=\frac{\delta S}{\delta A_t}=n, \\
j^z&=\frac{\delta S}{\delta A_z}=-\pi^{ z} \\
j^{\bar z}&=\frac{\delta S}{\delta A_{\bar z}}=-\pi^{\bar z}
\end{split}
\eeq
we express this Lagrangian in terms of the dual gauge fields $b=j^0$ and $e^i=\varepsilon^{i}_{\,\, j} j^j$
\beq
\begin{split}
\tilde{\mathcal{L}}_0&= b A_t-i\big(e^{\bar z}A_{\bar z}-e^z A_z \big)+\frac{1}{2b} \big( e^{\bar z} \partial_{\bar z} b+ e^z\partial_z b \big) \\
&=\varepsilon^{\mu\nu\rho} A_\mu \partial_{\nu} a_{\rho}+\frac{1}{2b} e^i \partial_i b.
\end{split}
\eeq
\subsection{Angular momentum of the LLL superfluid wave-function}
The single-particle $U(1)$ gauge-invariant  angular momentum operator is defined as
\beq
J_z=-i\varepsilon_i^{\, j} x^i D_j=(z D_z-\bar z D_{\bar z}),
\eeq
where $D_i=\partial_i - i A_i$.
The action of the operator $J_z$ on the LLL wave-function $\psi_{LLL}=f(z)e^{-|z|^2/(4 l_B^2)}$ is
\beq
J_z \psi_{LLL}=\Big[z \partial_z f- \frac B 2 |z|^2 f \Big]e^{-|z|^2/(4 l_B^2)},
\eeq
where we used the LLL condition $D_{\bar z} \psi_{LLL}=0$. Consider now the Gross-Pitaevskii LLL wave-function $\Psi_{LLL}$ which is normalized to the number of bosons $\int d^2 x \underbrace{|\Psi_{LLL}|^2}_{n}=N$. After integration by parts, we find the expectation value of the angular momentum $J_z$ in the Gross-Pitaevskii LLL state to be simply
\beq
\langle \Psi_{LLL}| J_z| \Psi_{LLL} \rangle =-N.
\eeq

Notice that in a constant magnetic field the operator $J_z$ does not generate rotations in the plane. The generator of rotations $L_z$ can be defined as
\beq
L_z=J_z+\frac 1 2 B x^2,
\eeq
which in the symmetric gauge reduces to
\beq
L_z=-i\varepsilon_i^{\, j} x^i \partial_j=(z \partial_z-\bar z \partial_{\bar z}). 
\eeq
The action of the operator $L_z$ on the Gross-Pitaevskii LLL wave-function is given by
\beq
L_z \Psi_{LLL}= e^{-|z|^2/(4 l_B^2)} z \partial_z f,
\eeq
 which after integration by parts results in the expectation value of  $L_z$ in this state to be
\beq
\langle \Psi_{LLL}| L_z| \Psi_{LLL} \rangle =\frac B 2 \int d^2 x |z|^2 n-N.
\eeq
Thus the angular momentum $L_z$ is given by the solid-body angular momentum $L^{sb}_z=I_{sb}\Omega=m \Omega \int d^2x n |z|^2$ shifted downwards by the number of bosons $N$.

\subsection{Comparison with  Ref.~\cite{Baym2003} }
Here we compare equations of motion derived from the Lagrangian \eqref{LqLLL} in the main text with the hydrodynamic equations derived by Baym in Ref.~\cite{Baym2003} for a rotating bosonic superfuid.
We first inroduce the rotation frequency $\Omega=B_0/2m$ and rewrite the quadratic Lagrangian \eqref{LqLLL} as
\beq
\begin{split}
\mathcal{ L}^{(2)}_{vc}
&=\underbrace{\frac{m}{2n_0}\mathbf{e}^2+\frac{1}{2n_0} e^i \partial_i \delta b}_{\text{NLO}}\underbrace{-\frac{m c_s^2}{2n_0}\delta b^2+2m \Omega e_i u^i -n_0m \Omega\epsilon_{ij}u^i\dot{u}^j-\mathcal{E}^{(2)}_{\text{el}}(\partial u) }_{\text{LO}}.
\end{split}
\eeq
This gives rise to the following equations of motion
\begin{subequations}\label{Bc1}
\begin{align}
&\partial_i (u^i+\frac{1}{2n_0 \Omega} e^i+\underline{\frac{1}{4m \Omega n_0} \partial^i \delta b})=0, \label{Bc1a} \\
& \dot e^i+ c_s^2 \epsilon^{ij}\partial_j \delta b+2 n_0 \Omega \dot u^i+\underline{\frac{1}{2m}\big( \partial^i \delta \dot b+\epsilon^{ik} \partial_k \partial_j e^j} \big)=0, \label{Bc1b} \\
& 2m \Omega e_i- 2 n_0 m \Omega \epsilon_{ij}\dot u^j+\partial_j \frac{\partial \mathcal{E}^{(2)}_{\text{el}}}{\partial \partial_j u^i}=0, \label{Bc1c}
\end{align}
\end{subequations}
where underlined terms originate from the Berry term.

If one ignores the underlined terms in
Eq.~(\ref{Bc1}), and identify now $e^i=\epsilon^{ij}j_j=\epsilon^{ij}
n_0 v_j$, these equations can be written as
\begin{align}
  & \bm\nabla\cdot\mathbf{u}
    + \frac1{2\Omega}\bm\nabla\times\mathbf{v} = 0, \label{Baym2} \\
  & m \partial_t\mathbf{j} + \bm{\nabla} P -2mn_0 \mathbf{\dot u}\times\bm\Omega = 0, \label{Baym8a}\\
  & 2mn_0(\mathbf{v}-\mathbf{\dot u})\times\bm\Omega -\bm\sigma = 0, \label{Baym10}
\end{align}
where $\sigma_i =\delta E^{(2)}_{\text{el}}/\delta u^i= - \partial_j (\partial
\mathcal{E}^{(2)}_{\text{el}}/\partial(\partial_j u^i))$.
Equations~(\ref{Baym2}) coincides with Eq.~(2) of Ref.~\cite{Baym2003}, 
and the curl of Eq.~(\ref{Baym10}) coincides with
Eq.~(10) of Ref.~\cite{Baym2003}.  A linear combination of
Eqs.~(\ref{Baym8a}) and (\ref{Baym10}),
\begin{equation}
  m\partial_t \mathbf{j} +\bm\nabla P - 2m \mathbf{j}\times\bm\Omega +
     \bm\sigma = 0
\end{equation}
is the equation of momentum conservation and coincides with Eq.~(8) of
Ref.~\cite{Baym2003}.

The agreement with Ref.~\cite{Baym2003} goes away, however, when the
underlined terms in Eq.~(\ref{Bc1}) are taken into account. We have
checked that an alternative identification of the velocity,
$e^i=\epsilon^{ij} n v_j- \partial_i n/2m$, does not help to
reduce the full version of Eq.~\eqref{Bc1} to the form of
Ref.~\cite{Baym2003} due to the last term on the second line in
Eq. \eqref{Bc1b}.  

Now we write down the full equations \eqref{Bc1} in the LLL limit,
taking $m\to0$, $\Omega\to\infty$, $c_s^2\to\infty$, $2m\Omega\equiv
B_0=\textrm{fixed}$, $mc_s^2=\textrm{fixed}$
\begin{subequations}
\begin{align}
  & \bm\nabla\cdot\mathbf{u} + \frac1{2 B_0 n_0}\nabla^2 n = 0, \label{hydro1}\\
  & \bm\nabla P -  \mathbf j \times \mathbf B_0 + \bm\sigma
  + \frac{n_0}2 \bm\nabla (\bm\nabla\cdot\mathbf v)\times \mathbf{\hat z}
  + \frac{n_0}2 \bm\nabla(\bm\nabla\times\mathbf v) = 0, \label{hydro2}\\
  & n_0 (\mathbf v - \mathbf{\dot u})\times\mathbf B_0 - \bm\sigma = 0. \label{hydro3}
\end{align}
\end{subequations}
In deriving Eq.~(\ref{hydro2}) we have combined Eqs.~(\ref{Bc1b}) and
(\ref{Bc1c}) and made use of the continuity equation $\dot n +
n_0\bm\nabla\cdot\mathbf v=0$.


\subsection{Transformation of stress tensor under Galilean boost in the LLL regime}
The energy-momentum tensor transforms under diffeomorphisms as
\beq
\delta T^{\mu\nu}=-\xi^\rho \partial_\rho T^{\mu\nu}+T^{\rho \nu}\partial_\rho \xi^\mu+T^{\mu \rho}\partial_\rho \xi^\nu.
\eeq
For the Galilean boost $\xi^i=\beta^i t$ we find that the stress tensor transforms as
\beq
\delta T^{ij}=-\beta^k t \partial_k T^{ij}+T^{0 j} \beta^i+T^{i0}\beta^j.
\eeq
The momentum density is given by
$
T^{0i}=m j^i-\text{g}\varepsilon^{ij}\nabla_j n/4
$\ which implies
\beq
\delta T^{ij}=-\beta^k t \partial_k T^{ij}+2 m j^{(i}\beta^{j)}-\frac{\text{g}}{2}\varepsilon^{(ik}\beta^{j)}\nabla_k n.
\eeq
In the LLL limit $m\to 0$ the second term in the transformation law disappears.

\subsection{Particle number linear response for general $\text{g}$}
In order to determine linear response of the particle current to an external $U(1)$ perturbation, we first solve linearized equations of motion in the presence of the  source $\mathcal{A}_\mu$. The quadratic part of the Lagrangian of an infinite vortex crystal is given by 
\beq \label{LqLLLS}
\begin{split}
\mathcal{ L}^{(2)}_{vc}&=\frac{m e^2}{2n_0}+\frac{1}{2n_0} e^i \partial_i \delta b-\frac{\varepsilon''}{2}\delta b^2-\frac{B_0 n_0}{2} \epsilon_{ij}u^i\dot{u}^j+B_0 e_i u^i-\mathcal{E}^{(2)}_{\text{el}}(\partial u)  +\frac{(\text{g}-2)}{4m} \mathcal{B} (n_0+\delta b) + \epsilon^{\mu\nu\rho}  \mathcal{A}_\mu \partial_\nu a_\rho, \\
\end{split}
\eeq
where $\mathcal{B}=\varepsilon^{ij}\partial_i \mathcal{A}_j$ and the elastic energy density
\beq \label{elen}
\begin{split}
\mathcal{E}^{(2)}_{\text{el}}(\partial u)&=2C_1 (\partial_i u^i)^2+ C_2 \big[ (\partial_x u^x- \partial_y u^y)^2+(\partial_y u^x+\partial_x u^y)^2\big], 
\end{split}
\eeq
where $C_1$ and $C_2$ define the compressional and shear moduli of the vortex crystal.Throughout the computation of the conductivity we will keep the mass $m$ finite and will take the LLL limit $m\to 0$ in the very end.

The Gauss law extracted from the Lagrangian \eqref{LqLLLS} is
\beq \label{GLLLLSg}
\partial_i \big(m e^i+\frac 1 2 \partial^i \delta b+n_0 B_0 u^i \big)= -n_0 \mathcal{B}.
\eeq
To discuss the remaining equations of motion, we will work in the temporal gauge $a_t=0$ and restrict our attention only to sources and solutions that have the form of a plane wave propagating along $x$ axis. 
The equation of motion in Fourier space ($k=-i \partial_x$, $\omega=i\partial_t$) are
\beq \label{EOMLLLSg}
\left(
\begin{array}{cccc}
 -m \omega^2 & \frac{i}{2}  k^2 \omega  & -i B_0
   \omega  n_0 & 0 \\
 -\frac{i}{2}  k^2 \omega  &
   -m \omega^2+\epsilon'' k^2 n_0 & 0 & -i B_0
   \omega  n_0 \\
 -\frac{i B_0 \omega }{2} & 0 & -k^2
   \left(2 C_1+C_2\right) &
   \frac{i}{2}  B_0 \omega  n_0 \\
 0 & -\frac{i B_0 \omega }{2} &
   -\frac{i}{2}  B_0 \omega  n_0 & -k^2
   C_2 \\
\end{array}
\right) 
\left( \begin{array}{c} a_x \\ a_y \\ u_x \\ u_y \end{array}  \right)=
\left( \begin{array}{c} i n_0 \omega \mathcal{A}_y   \\
 -i n_0(\omega \mathcal{A}_x+ k \mathcal{A}_t)+\frac{\text{g}-2}{4m} n_0 k^2 \mathcal{A}_y   \\ 0 \\ 0 \end{array}  \right).
\eeq
We now substitute the solution of the last equation into the expression for the $U(1)$ current 
\beq
j^i=\frac{\delta S}{\delta A_i}=-\varepsilon^{ij} e_j+\frac{\text{g}-2}{4m}\varepsilon^{ij} \partial_j b
\eeq
and extract the conductivity tensor from the definition
\beq \label{LRd}
\begin{split}
\sigma^{ij}(\omega, \mathbf{k})&=-\frac{\delta j^i}{\delta E_j}=\frac{\delta j^i}{i \omega \delta A_j}=\frac{\delta j^i}{i k_j \delta A_t}.
\end{split}
\eeq
We perform this computation using the software package Mathematica and find the Hall conductivity $\sigma^H=(\sigma^{xy}-\sigma^{yx})/2$
\beq \label{Hcg1}
\begin{split}
\sigma_H&=\frac{n_0}{B_0}+\frac{(\text{g}-4) n_0 }{4 B_0^2}k^2+O(k^4)\,
\end{split}
\eeq
and the longitudinal conductivity $\sigma_L= (\sigma^{xx}+\sigma^{yy})/2$ whose singular part in the limit $m\to 0$ is
\beq \label{Hcg2}
\begin{split}
\sigma^{sing}_L&=-\frac{i(\text{g}-2)n_0}{4B_0 \omega m}k^2+O(k^4)\,.
\end{split}
\eeq
Galilean invariance implies that the AC Hall viscosity can be completely determined by the conductivity tensor provided one knows the momentum expansion of the latter up to quadratic order. For $\text{g}=0$ such a relation was derived in Ref.~\cite{Bradlyn2012} [see Eq.~(4.14) in that paper] and later generalized to an arbitrary value of $\text{g}$ in Ref.~\cite{Geracie2015} [see Eq.~(70) in that paper]
\beq
\eta_{H}=\frac{1}{2} \partial^2_{k_x} \Big[(m^2\omega^2+B_0^2) \sigma_H(\mathbf{k})-2i m \omega B_0 \sigma_{L}(\mathbf{k}) \Big]+\frac{\text{g}}{4} \Big[ B_0 \sigma_H(\mathbf{k}=0)-i m \omega \sigma_L(\mathbf{k}=0) \Big]
\eeq
Taking now the limit $m\to 0$ and using Eqs. \eqref{Hcg1} and \eqref{Hcg2}, we find $\eta_H=0$ for an arbitrary value of the gyromagnetic factor $\text{g}$.

\end{widetext}

\end{document}